\documentclass[reprint, preprintnumbers, aps, prl, floatfix, superscriptaddress, nofootinbib]{revtex4-2}

\usepackage{silence}
\usepackage{amssymb, amsmath, physics, graphicx, xcolor, orcidlink, fontawesome, hyperref, bm}

\input{custom_I_symbol}

\newcommand{\IoA}{Institute of Astronomy, University of Cambridge, Madingley Road, Cambridge, CB3 0HA, UK}
\newcommand{\KICC}{Kavli Institute for Cosmology, University of Cambridge, Madingley Road, Cambridge, CB3 0HA, UK}
\newcommand{\DAMTP}{Department of Applied Mathematics and Theoretical Physics, Centre for Mathematical Sciences, University of Cambridge, Wilberforce Road, Cambridge, CB3 0WA, UK}
\newcommand{\SJC}{St John's College, Cambridge, CB2 1TP, United Kingdom}

\begin{document}

\title{Black Hole Fine Structure}

\author{Adrian Ka-Wai Chung\,\orcidlink{0000-0003-2020-3254}}
\email{kwc43@cam.ac.uk}
\affiliation{\SJC} \affiliation{\DAMTP}

\author{Richard Dyer\,\orcidlink{0009-0008-3720-6092}}
\email{richard.dyer@ast.cam.ac.uk}
\affiliation{\IoA}

\author{Tomasz Kinowski\,\orcidlink{0009-0004-9474-7654}}
\email{tk593@cam.ac.uk}
\affiliation{\IoA} \affiliation{\KICC}

\author{Christopher J.\ Moore\,\orcidlink{0000-0002-2527-0213}}
\email{cjm96@cam.ac.uk}
\affiliation{\IoA} \affiliation{\DAMTP} \affiliation{\KICC}

\date{\today}

\begin{abstract}
    In General Relativity, axial and polar black-hole perturbations share identical quasinormal-mode spectra, a property known as isospectrality.
    New physics can break this degeneracy, producing a spectral splitting analogous to atomic fine structure.
    We perform a theory-agnostic search for such splitting in gravitational-wave observations.
    We find no evidence for isospectrality breaking and place a combined 90\% upper limit of $<28.5$\% on the relative frequency splitting. 
    We map these constraints onto several modified-gravity theories, constraining the coupling-lengths to be $\lesssim 28\,\mathrm{km}$.
\end{abstract}

\maketitle

{\textit{\textbf{Introduction}}---}The final stage of a black-hole (BH) merger, known as the ringdown, provides a powerful test of General Relativity (GR)
\cite{LIGOScientific:2020tif,LIGOScientific:2021sio,LIGOScientific:2025slb,LIGOScientific:2025wao,LIGOScientific:2026wpt,LIGOScientific:2026oim, Berti:2015itd, Cardoso:2019rvt, Dreyer:2003bv}.
Besides probing some of the strongest and most dynamical gravitational fields in nature, the ringdown is amenable to a semi-analytic description through BH perturbation theory, making its physical interpretation particularly transparent.

The ringdown gravitational-wave (GW) signal can be modeled using a superposition of quasinormal modes (QNMs) \cite{Vishveshwara:1970zz, Press:1971wr, Davis:1971gg, Berti:2005ys}, damped oscillations whose complex frequencies are determined by the mass and spin of the remnant BH in GR.
Measuring these frequencies is the basis of BH spectroscopy \cite{Dreyer:2003bv, Berti:2005ys, Berti:2016lat}, named by analogy with atomic spectroscopy.
Additional gravitational interactions can modify this spectrum \cite{Cano:2023jbk,Chung:2024ira,Chung:2024vaf,Chung:2025gyg,Li:2025fci,Li:2023ulk,Weller:2026jqu} thereby providing signatures of unknown physics.

A remarkable property of vacuum BHs in GR is isospectrality \cite{Chandrasekhar:1975zza, Chandrasekhar:1985kt}, meaning that axial (odd-parity) and polar (even-parity) gravitational perturbations share identical QNM spectra.
This degeneracy is generically broken in non-vacuum situations and in many modified theories of gravity
\cite{Cano:2023jbk,Chung:2024ira,Chung:2024vaf,Chung:2025gyg,Li:2025fci,Li:2023ulk,Weller:2026jqu},
giving rise to distinct QNM branches with different frequencies and damping times.
In the language of atomic spectroscopy, this lifting of the degeneracy and splitting in the QNM spectrum can be thought of as fine structure.
Observing such fine structure would provide evidence for new physics, possibly including a departure from GR, although some modified theories nevertheless preserve isospectrality \cite{Weller:2026jqu}. 

In this Letter, we perform a theory-agnostic search for isospectrality breaking using the full inspiral--merger--ringdown signals of 13 selected binary-BH mergers from GWTC-5.0 \cite{LIGOScientific:2026sit}.
Previously, isospectrality in specific gravity theories has been tested in ringdown-only analyses \cite{Maenaut:2024oci,Chung:2025ucb}. 
We instead introduce independent frequency and damping-time shifts and splittings as phenomenological observables and constrain them directly from the data.
Except for GW231028\_153006, a known outlying event \cite{LIGOScientific:2025slb}, we find no evidence for new physics. 
This exceptional event shows evidence for a shift, but not a split, in the QNM frequencies. 
Overall, we find no evidence for any QNM splitting associated with isospectrality breaking and BH fine structure. 

Assuming common shift and splitting parameters across events, we also combine the event-level posteriors to obtain catalog-level constraints.
We then map these theory-agnostic constraints onto the coupling length scales of several representative effective-field-theory (EFT) extensions of GR that predict broken isospectrality.
In this way, a single phenomenological search for BH fine structure is used to test a broad class of theories beyond GR.

{\textit{\textbf{Modeling broken isospectrality}}---}When isospectrality is broken, a QNM with indices $(l,m,n)$ splits into two branches with different complex frequencies; in analogy with atomic fine structure, there is a shift and a split in the complex QNM frequencies.
We adopt a minimal phenomenological description in which the two branches share the same inspiral--merger dynamics, but acquire different QNM frequencies and damping times through the transition into ringdown.
The observed signal is therefore modeled as a superposition of two otherwise identical inspiral--merger--ringdown waveforms that differ only in the complex frequencies of their fundamental QNMs.
Guided by this picture, we write
\begin{equation} \label{eq:waveform_model}
    \begin{split}
        h(t;\bm{\Theta}) ={}&
        (1-\delta A)\,
        \underline{h}\!\left(
        t;\bm{\Theta},\delta f^{(1)},\delta\tau^{(1)}
        \right)
        \\
        &+
        \delta A\,
        \underline{h}\!\left(
        t;\bm{\Theta},\delta f^{(2)},\delta\tau^{(2)}
        \right),
    \end{split}
\end{equation}
where $\bm{\Theta}$ denotes the GR source parameters, $\delta A\in[0,0.5]$ controls the relative amplitude of the two branches
\footnote{More generally, the relative excitation coefficient of the two branches may be complex. 
Here we assume that the two branches have the same excitation phase in the GR limit, so that the leading-order mixing coefficient is real.
Within this prescription, coupling-induced complex corrections enter only beyond leading order; see the Supplemental Material.}.
Here $\underline{h}(t;\bm{\Theta},\delta f,\delta\tau)$ denotes a generic inspiral--merger--ringdown waveform with a single complex fundamental QNM frequency $\omega_{220}=2\pi f-i/\tau$, where 
\begin{align}
    f = f^{\rm GR}(1+\delta f) \quad \mathrm{and} \quad \tau = \tau^{\rm GR}(1+\delta \tau), 
\end{align}
and the superscript GR denotes the Kerr prediction and the $\delta$-values are relative fractional deviations.
Hereafter we drop the QNM indices from our notation for simplicity and always refer to the fundamental, prograde $(2,2,0)$ mode.

\begin{figure*}[htp!]
    \centering
    \includegraphics[width=\linewidth]{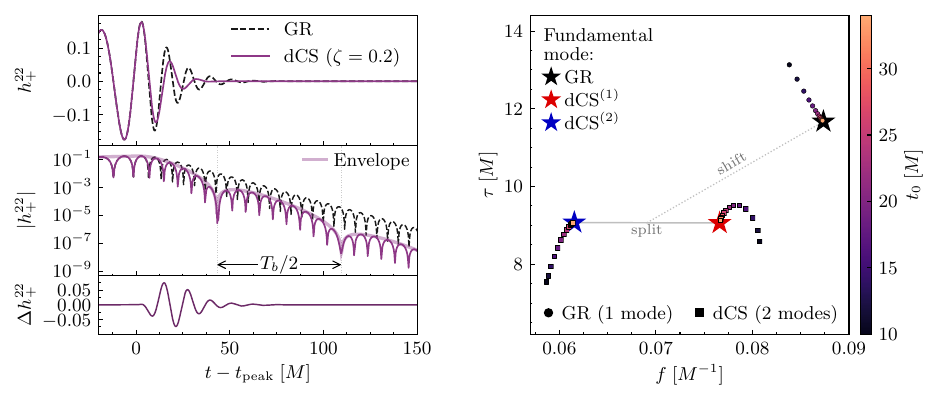}
    \caption{ \label{fig:01}
        A plot of our phenomenological waveform for isospectrality breaking showing the unmodified inspiral and the presence of two fundamental QNMs with shift and a split in their complex frequencies relative to Kerr. 
        Observe the beating pattern with period $T_b$ due to the frequency split.
        \emph{Left:} Plus polarization of the $(l,m)=(2,2)$ waveform showing the parity-dependent QNM shifts relative to the usual (GR) waveform. 
        The example shifts plotted here are for dynamical Chern--Simons (dCS) gravity with $\zeta\equiv(\ell/M)^4=0.2$ and $\delta A=0.5$.
        The waveform is plotted on linear (top) and logarithmic (middle) scales and the residuals relative to GR waveform are shown at the bottom. 
        \emph{Right:} Stars show the single fundamental (2,2,0) complex QNM frequency for a Kerr BH (GR) and the two parity-dependent frequencies for a dCS BH. 
        The colored points show the results of a nonlinear least-squares fit to the waveform for the frequency as a function of start time, $t_0$; one damped sinusoid was used for the GR fit (circles) and two for the dCS fit (squares) which show the fundamental QNM frequencies present in the ringdown.
    }
\end{figure*}

\begin{table*}[tp!]
    \centering

    \begingroup
    \renewcommand{\arraystretch}{1.45}

    \resizebox{0.98\textwidth}{!}{%
        \begin{tabular}{lcccccccccccr}
    \hline\hline
    Run & Event name & SNR
    & $\delta A$
    & $\delta f^{(1)}$
    & $\delta \tau^{(1)}$
    & $\delta f^{(2)}$
    & $\delta \tau^{(2)}$
    & $\Delta_f$
    & $\Delta_{\tau}$
    & $\log \mathcal{B}^{\barredI}_{I}$ \\
    \hline
        O1  & GW150914         & $26.0^{+0.1}_{-0.2}$ & 
        $0.37^{+0.12}_{-0.22}$ & $\phantom{+}0.062^{+0.292}_{-0.080}$ & $\phantom{+}0.262^{+0.497}_{-0.468}$ & $\phantom{+}0.171^{+0.225}_{-0.252}$ & $\phantom{+}0.452^{+0.484}_{-0.793}$ & $\leq$ 0.326 & $\leq$ 0.874 & $-3.37$ \\

        O3a & GW190521\_074359 & $25.0^{+0.1}_{-0.2}$ & 
        $0.32^{+0.15}_{-0.20}$ & $\phantom{+}0.049^{+0.111}_{-0.086}$ & $\phantom{+}0.240^{+0.413}_{-0.506}$ & $\phantom{+}0.109^{+0.434}_{-0.375}$ & $-0.217^{+0.845}_{-0.538}$ & 
        $\leq$ 0.423 & $\leq$ 1.14 & $-3.10$ \\

        O3b & GW200129\_065458 & $26.8^{+0.2}_{-0.2}$ & 
            $0.29^{+0.18}_{-0.20}$ & 
            $\phantom{+}0.001^{+0.060}_{-0.080}$ & $\phantom{+}0.492^{+0.414}_{-0.498}$ & 
            $-0.074^{+0.515}_{-0.313}$ & $\phantom{+}0.192^{+0.699}_{-0.742}$ & $\leq0.409$ & $\leq0.514$ & $-3.49$ \\

            & GW200224\_222234 & $20.0^{+0.2}_{-0.2}$ & 
            $0.30^{+0.17}_{-0.21}$ & $\phantom{+}0.054^{+0.136}_{-0.131}$ & $\phantom{+}0.372^{+0.502}_{-0.536}$ & $\phantom{+}0.134^{+0.462}_{-0.358}$ & $\phantom{+}0.150^{+0.725}_{-0.713}$ & 
            $\leq$ 0.457 & $\leq$ 1.02 & $-2.58$ \\

        O4a & GW230927\_153832 & $19.7^{+0.2}_{-0.2}$ & 
        $0.27^{+0.18}_{-0.19}$ & 
        $-0.024^{+0.085}_{-0.076}$ & $\phantom{+}0.554^{+0.370}_{-0.558}$ & $-0.103^{+0.694}_{-0.552}$ & $\phantom{+}0.177^{+0.707}_{-0.766}$ & 
        $\leq$ 0.620 & $\leq$ 1.13 & $-1.91$ \\

            & GW231028\_153006 & $21.0^{+0.2}_{-0.2}$ & 
            $0.35^{+0.13}_{-0.22}$ & $-0.367^{+0.090}_{-0.105}$ & $\phantom{+}0.143^{+0.538}_{-0.377}$ & $-0.177^{+0.409}_{-0.478}$ & $-0.400^{+1.040}_{-0.356}$ & 
            $\leq$ 0.509 & $\leq$ 1.02 & +3.27 \\

            & GW231206\_233901 & $21.0^{+0.1}_{-0.2}$ & 
            $0.38^{+0.11}_{-0.22}$ & $\phantom{+}0.110^{+0.181}_{-0.070}$ & $\phantom{+}0.498^{+0.429}_{-0.881}$ & $\phantom{+}0.037^{+0.751}_{-0.504}$ & $-0.557^{+1.320}_{-0.228}$ & 
            $\leq$ 0.704 & $\leq$ 1.51 & $-2.25$ \\

            & GW231226\_101520 & $33.7^{+0.1}_{-0.1}$ & 
            $0.28^{+0.16}_{-0.15}$ & $\phantom{+}0.058^{+0.040}_{-0.044}$ & $\phantom{+}0.257^{+0.274}_{-0.284}$ & $\phantom{+}0.419^{+0.102}_{-0.404}$ & $-0.035^{+0.727}_{-0.427}$ & 
            $\leq$ 0.434 & $\leq$ 0.795 & $-3.48$ \\

        O4b & GW240615\_113620 & $23.8^{+0.1}_{-0.2}$ & 
        $0.40^{+0.09}_{-0.23}$ & $-0.110^{+0.122}_{-0.111}$ & $\phantom{+}0.471^{+0.401}_{-0.429}$ & $-0.078^{+0.120}_{-0.180}$ & $\phantom{+}0.642^{+0.316}_{-0.652}$ & 
        $\leq$ 0.209 & $\leq$ 0.650 & $-0.728$ \\

            & GW240621\_195059 & $27.6^{+0.1}_{-0.2}$ & 
            $0.28^{+0.19}_{-0.19}$ & $-0.012^{+0.075}_{-0.062}$ & $\phantom{+}0.498^{+0.370}_{-0.455}$ & $-0.048^{+0.632}_{-0.500}$ & $-0.274^{+1.010}_{-0.478}$ & 
            $\leq$ 0.543 & $\leq$ 1.32 & $-2.86$ \\

            & GW240920\_124024 & $37.0^{+0.1}_{-0.1}$ & 
            $0.34^{+0.14}_{-0.20}$ & $\phantom{+}0.040^{+0.049}_{-0.038}$ & $\phantom{+}0.413^{+0.369}_{-0.490}$ & 
            $\phantom{+}0.365^{+0.125}_{-0.375}$ & $-0.103^{+0.744}_{-0.348}$ & 
            $\leq$ 0.430 & $\leq$ 0.984 & $-1.58$ \\

            & GW241127\_061008 & $31.2^{+0.1}_{-0.2}$ & 
            $0.39^{+0.10}_{-0.19}$ & $\phantom{+}0.195^{+0.444}_{-0.208}$ & $-0.335^{+0.507}_{-0.306}$ & $\phantom{+}0.115^{+0.670}_{-0.150}$ & $-0.018^{+0.622}_{-0.589}$ & 
            $\leq$ 0.675 & $\leq$ 0.956 & $-0.24$ \\

            & GW250114\_082203 & $76.9^{+0.0}_{-0.1}$ & 
            $0.27^{+0.20}_{-0.19}$ & $\phantom{+}0.025^{+0.027}_{-0.033}$ & $\phantom{+}0.110^{+0.227}_{-0.284}$ & $\phantom{+}0.055^{+0.868}_{-0.165}$ & $-0.213^{+0.643}_{-0.389}$ & 
            $\leq$ 0.795 & $\leq$ 0.778 & $-5.41$ \\
    \hline\hline
\end{tabular}%
    }
    \endgroup
    \caption{ \label{tab:GW_events}
        Individual event posteriors for the 13 selected GW events. 
        All deformation parameters refer to the fundamental $(\ell,m,n)=(2,2,0)$ QNM and superscripts (1) and (2) label the two QNM branches. 
        The SNR column gives the reported median network SNR and its 90\% credible interval from Refs.~\cite{LIGOScientific:2026wpt, LIGOScientific:2026oim}. 
        For $\delta A$, the signed QNM shifts, and the common-mode parameters, we quote the posterior median and equal-tailed 90\% credible interval. 
        For $\Delta_f$, and $\Delta_\tau$, we quote one-sided 90\% credible upper limits measured from zero. 
        The last column reports the natural logarithm of the Bayes factor, comparing the isospectrality-breaking hypothesis $H_{\barredI}$ with the isospectral GR hypothesis $H_I$; negative values favor $H_I$. 
    }
\end{table*}

Because the two branches have identical GR source parameters and their weights sum to unity, they reproduce the same inspiral--merger signal and differ only in the ringdown.
The two branches of the waveform are denoted by superscripts $(1)$ and $(2)$ which refer to the louder and quieter branches respectively;
observationally, we do not know which of the axial and polar perturbations is louder in the observed signal, and imposing $\delta A<1/2$ avoids any label-switching degeneracy.
Relative to the underlying GR waveform, the construction introduces five additional parameters: $\delta A$, $\delta f^{(1)}$, $\delta\tau^{(1)}$, $\delta f^{(2)}$, and $\delta\tau^{(2)}$.

This construction relies on a waveform model where the QNM frequencies and damping times can be varied consistently through the merger--ringdown transition.
We use \texttt{pSEOBNRv5PHM} for $\underline{h}$ \cite{Pompili:2025cdc,Mihaylov:2023bkc}; a multipolar,
spin-precessing effective-one-body waveform model for binary BHs in quasicircular orbits.
Example waveforms are plotted in Fig.~\ref{fig:01}.
Further details of the waveform implementation, parameter priors, and an injection study into actual detector noise are given in the Supplemental Material.

{\textit{\textbf{Event selection}}---}Following the event-selection criteria adopted for the \texttt{pSEOBNRv5PHM} analysis in Ref.~\cite{LIGOScientific:2026oim}, we consider binary-BH events in GWTC-5.0 for which the signal-to-noise ratio (SNR) in both the pre- and post-inspiral portions of the signal are at least 8, to ensure that our results are not obscured by the degeneracies between the total mass and QNM frequency parameters.
We additionally require a total network SNR of at least 18, restricting the analysis to the loudest and most informative events. 
These criteria leave the 13 events listed in Tab.~\ref{tab:GW_events}.

{\textit{\textbf{Theory-agnostic results}}---} 
We perform Bayesian inference with this model comparing two hypotheses: an isospectral hypothesis, i.e.\ GR, $H_I$ described by parameters $\bm \Theta$ and an isospectral-breaking hypothesis, $H_{\barredI}$, in which the fundamental QNM splits into two branches with independent complex frequencies given by the five additional parameters described above.
The QNM frequencies in the $H_{\barredI}$ hypothesis have both a shift and a split.
The details of the Bayesian inference are described in the Supplemental Material.

For each event, we infer the marginalized posterior distributions of $\delta A$, $\delta f^{(1,2)}$, and $\delta\tau^{(1,2)}$.
The individual frequency and damping-time shifts contain both a common displacement of the two branches and a branch-to-branch splitting (see Fig.~\ref{fig:01}).
The former can be partially degenerate with changes in the inferred remnant mass and spin and therefore does not by itself constitute a clean observable of isospectrality breaking.
We focus instead on the following isospectrality-breaking frequency-splitting combinations
\begin{equation} \label{eq:Delta}
    \Delta_f = \left|\delta f^{(1)}-\delta f^{(2)}\right|,
    \quad
    \Delta_\tau = \left|\delta\tau^{(1)}-\delta\tau^{(2)}\right|.
\end{equation}
We also quantify the evidence for new physics in this analysis using the Bayes factor $\mathcal{B}^{\barredI}_{I}=p(d|H_{\barredI})/p(d|H_I)$.

\begin{figure}[tp!]
    \centering
    \includegraphics[width=\linewidth]{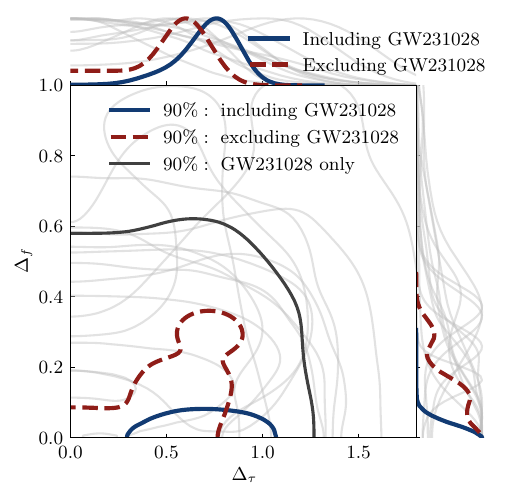}
    \caption{ \label{fig:catalog_PDF}
        The 90\% credible contours of the joint posterior distribution of the frequency and damping-time splittings, $\Delta_{f}$ and $\Delta_{\tau}$ of the (2,2,0) QNM, together with their corresponding one-dimensional marginalized posterior distributions. 
        The light-gray curves show the posteriors for the individual events. 
        The solid dark-blue curves show the combined posterior obtained using all 13 events, including GW231028\_153006, while the
        dashed dark-red curves show the combined posterior obtained after excluding GW231028\_153006. 
        The dark-gray contour in the central panel shows the 90\% credible region obtained from GW231028\_153006 alone.
    }
\end{figure}

Table \ref{tab:GW_events} lists the one-sided 90\% credible upper limits on $\Delta_{f}$, and $\Delta_{\tau}$, together with the posterior medians and equal-tailed 90\% credible intervals of $\delta A$, $\delta f^{(1,2)}$ and $\delta\tau^{(1,2)}$ and the corresponding values of $\log\mathcal{B}^{\barredI}_{I}$.
Most individual frequency and damping-time shifts are consistent with zero at 90\% credibility, although a small number of event-level posteriors show modest offsets.
These offsets do not, however, constitute evidence for isospectrality breaking, since all, except GW231028\_153006, yield Bayes factors favoring $H_I$ ($\log\mathcal{B}^{\barredI}_{I} < 0 $). 
GW231028\_153006 is the only analyzed event of positive $\log\mathcal{B}^{\barredI}_{I}$. 
However, this does not indicate detection of isospectrality breaking, as this event is known to exhibit significant waveform-dependent systematic variations \cite{LIGOScientific:2025slb}.
Fig.~\ref{fig:catalog_PDF} shows the individual and combined posteriors on $\Delta_{f}$ and $\Delta_{ \tau}$.
Under the additional assumption that the QNM splittings are consistent with zero across the population, we combine the 13 event-level posteriors.
The marginalized frequency-splitting posterior is consistent with $\Delta_{f}=0$ (with 90\% upper credible interval from zero being at $\Delta_{f} = 0.285$), whereas the damping-time splitting peaks away from zero.
Similar displacements of the damping-time posterior away from the GR value have previously been observed in BH spectroscopy \cite{LIGOScientific:2021sio,LIGOScientific:2026wpt,LIGOScientific:2026oim}.
The latter is not accompanied by Bayesian evidence for non-isospectrality and is therefore not interpreted as evidence for isospectrality breaking. 

\begin{figure}[t]
    \centering
    \includegraphics[width=\linewidth]{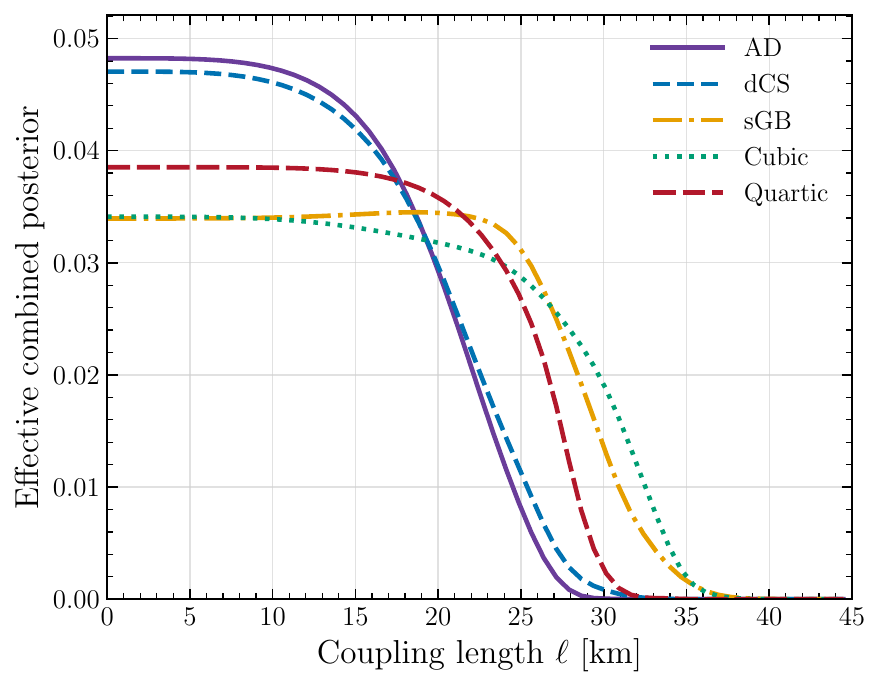}
    \caption{ \label{fig:PDF_ell}
        The effective posterior distribution of the coupling length scales of various beyond-GR extensions, derived from combining the marginalized posterior of $\delta f^{(1,2)}$ and $\delta \tau^{(1,2)}$ across all events (excluding GW231028\_153006). 
    }
\end{figure}

{\textit{\textbf{Fundamental-physics implications}}---}The theory-agnostic constraints on isospectrality breaking can be
translated into bounds on specific modifications of GR once their parity-dependent QNM spectra are specified.
As representative EFTs, we consider axi-dilaton (AD) \cite{Cano:2021rey, Kallosh:2022vha}, dynamical Chern--Simons (dCS) \cite{Alexander:2009tp}, scalar Gauss--Bonnet (sGB), cubic \cite{Bueno:2016xff}, and quartic gravity \cite{Endlich:2017tqa}, spanning several classes of higher-curvature corrections to the Einstein--Hilbert action \cite{Endlich:2017tqa, Cano:2019ore}.
In these theories, isospectrality is broken and the QNM spectrum acquires parity-dependent corrections that can be written schematically as
\begin{equation} \label{eq:qg_qnm_expansion}
    M\omega_{\rm (X)}(\ell,a)=M\omega^{\rm GR}(a)+\left(\frac{\ell}{M}\right)^k \Omega_{\rm (X)}(a),
\end{equation}
where $X\in\{{\rm A},{\rm P}\}$ labels the axial and polar sectors, $\omega^{\rm GR}$ is the parity-independent GR QNM frequency, $M$ and $a$ are the source-frame remnant mass and dimensionless spin, $\ell$ is the coupling length scale, $k$ is the EFT order of the modification to GR action (see Tab.~\ref{tab:theory_constraints}), and $\Omega_{\rm (X)}(a)$ encodes the leading-order parity-dependent correction.
We evaluate $\Omega_{\rm (X)}(a)$ using the QNM calculations of Refs.~\cite{Chung:2024ira,Chung:2024vaf,Chung:2025gyg, Chung:2025ucb,Cano:2023jbk}; the corresponding definitions of $\ell$ are given in the Supplemental Material.

Equation \eqref{eq:qg_qnm_expansion} predicts the frequency and damping-time splittings for any given $\ell$, remnant mass, and spin.
We therefore project the joint phenomenological posterior onto the QNM spectrum predicted by each theory, marginalizing over the remnant parameters and over the two possible assignments of the phenomenological branches (1 and 2) to the two parity sectors (A and P).
The full mapping procedure is described in the Supplemental Material.

Figure~\ref{fig:PDF_ell} shows the resulting effective posteriors on $\ell$, obtained by combining the 12 selected signals.
For all theories considered, the posteriors are consistent with the GR limit, $\ell=0$, in agreement with the absence of evidence for isospectrality breaking.
We obtain one-sided 90\% upper limits on $\ell$ of approximately $20$--$28\,{\rm km}$; the individual bounds are summarized in Tab.~\ref{tab:theory_constraints}.

\begin{table}[t]
    \begin{ruledtabular}
        \begin{tabular}{lcc}
            Theory
            & $k$
            & $\ell$ [km] \\
            \hline
            Axi--dilaton (AD)
            & 4
            & $\leq$ 20.3 \\
            
            Dynamical Chern--Simons (dCS)
            & 4
            & $\leq$ 20.9\\
    
            Scalar Gauss--Bonnet (sGB)
            & 4
            & $\leq$ 26.7\\
    
            Cubic gravity
            & 4
            & $\leq$ 27.9\\
    
            Quartic gravity
            & 6
            & $\leq$ 24.2\\
        \end{tabular}
    \end{ruledtabular}
    \caption{ \label{tab:theory_constraints}
        Combined constraints on the coupling length scales of the five beyond-GR theories considered in this work. 
        Here $k$ is the theory-dependent exponent in Eq.~\eqref{eq:qg_qnm_expansion}.
        The quoted bounds are one-sided 90\% credible upper limits on $\ell$.
    }
\end{table}

%
%
%
%

{\textit{\textbf{Discussion}}---}We have performed the first theory-agnostic search for the breaking of BH isospectrality using GW detections. 
We analyze 13 binary-BH signals, finding no robust evidence for BH fine structure. 
One event, GW231028\_153006, yields a Bayes factor that disfavors GR but does not show signs of isospectrality breaking, as this event is known to exhibit significant waveform-dependent systematics.
Excluding this event, the remaining 12 signals are consistent with isospectrality. 
Their combination places substantially tighter constraints on the frequency and damping-time splittings of the $(2,2,0)$ QNM. 
Although the combined damping-time splitting peaks away from zero, this behavior is not accompanied by evidence favoring the non-isospectral model and is therefore not interpreted as evidence for isospectrality breaking.

The absence of a detected splitting can nevertheless be used to probe fundamental physics. 
By mapping the theory-agnostic constraints onto the parity-dependent QNM spectra predicted by several modified-gravity theories, we obtain 90\% upper limits of $\lesssim 28\,{\rm km}$ on their coupling length scales. 
More broadly, the framework introduced here turns the lifting of BH isospectrality into an observable in its own right: GW data can first determine whether the QNM spectrum exhibits a ``fine structure'', independently of its theoretical origin, before that measurement is interpreted within particular theories of gravity. 
With louder ringdowns and improved QNM calculations, this approach will provide increasingly stringent and broadly applicable tests of the gravitational dynamics of BHs.

{\textit{\textbf{Acknowledgments}}---}
The authors thank Daniela Cors, Ulrich Sperhake and Seppe Staelens for insightful discussions. 
RD acknowledges support from the STFC for their PhD (Grant No.\ ST/Y509139/1) and A.K.W.C. acknowledges the Herchel Smith Research Fellowship and Title-E Fellowship at St John's College, Cambridge. 
TK acknowledges support for their PhD from the Science and Technology Facilities Council (STFC) through a Cambridge Astronomy Doctoral Landscape Award studentship (Grant No.\ 1732).
The calculations and results reported in this paper were produced using the computational resources of the Illinois Campus Cluster, which is a computing resource that is operated by the Illinois Campus Cluster Program (ICCP) in conjunction with National Center for Supercomputing Applications (NCSA), and the computational resources of Scientific and High Performance Computing at DAMTP, Cambridge.
This work used the large language model (LLM) \texttt{ChatGPT 5.6 Sol Extra High} to assist with literature review, coding, running scripts, documentation, and refining parts of this text, all closely supervised by A.K.W.C.
This material is based upon work supported by NSF's LIGO Laboratory which is a major facility fully funded by the National Science Foundation.
This research has made use of data or software obtained from the Gravitational Wave Open Science Center (gwosc.org), a service of the LIGO Scientific Collaboration, the Virgo Collaboration, and KAGRA. This material is based upon work supported by NSF's LIGO Laboratory which is a major facility fully funded by the National Science Foundation, as well as the Science and Technology Facilities Council (STFC) of the United Kingdom, the Max-Planck-Society (MPS), and the State of Niedersachsen/Germany for support of the construction of Advanced LIGO and construction and operation of the GEO600 detector. Additional support for Advanced LIGO was provided by the Australian Research Council. Virgo is funded, through the European Gravitational Observatory (EGO), by the French Centre National de Recherche Scientifique (CNRS), the Italian Istituto Nazionale di Fisica Nucleare (INFN) and the Dutch Nikhef, with contributions by institutions from Belgium, Germany, Greece, Hungary, Ireland, Japan, Monaco, Poland, Portugal, Spain. KAGRA is supported by Ministry of Education, Culture, Sports, Science and Technology (MEXT), Japan Society for the Promotion of Science (JSPS) in Japan; National Research Foundation (NRF) and Ministry of Science and ICT (MSIT) in Korea; Academia Sinica (AS) and National Science and Technology Council (NSTC) in Taiwan.

{\textit{\textbf{Data Availability}}---}This work analyzes GW strain data that are publicly available through the Gravitational Wave Open Science Center (GWOSC) at \url{https://gwosc.org/}. 
The waveform model developed in this work and code to perform the Bayesian inference are available in the git repository \cite{isospectrality_code}. 
The posterior samples of data analyses is available at \cite{isospectrality_data}.

\vspace{1cm}
\bibliographystyle{apsrev4-1}
\bibliography{references}

\clearpage
\onecolumngrid
\newpage
\begin{center}
  \textbf{\large{Supplemental Material}} \\
\end{center}
\twocolumngrid

\setcounter{equation}{0}
\setcounter{figure}{0}
\setcounter{table}{0}
\setcounter{page}{1}
\makeatletter
\renewcommand{\theequation}{S\arabic{equation}}
\renewcommand{\thefigure}{S\arabic{figure}}
\renewcommand{\thetable}{S\arabic{table}}

\section{Waveform model} \label{sec:model}

Gravitational perturbations of the $(\ell, m, n)$ QNM around a BH can be written as the sum of the two parts, according to their parities, 
\begin{align}
h_{\mu \nu} = h^{\rm (A)}_{\mu \nu} + h^{\rm (P)}_{\mu \nu}, 
\end{align}
where $h^{\rm (A)}_{\mu \nu}$ and $h^{\rm (P)}_{\mu \nu}$ are the axial and polar sectors of GWs respectively. 
These sectors are classified according to their properties with respect to the parity transformation. 
Mathematically, $h^{\rm (A)}_{\mu \nu}$ and $h^{\rm (P)}_{\mu \nu}$ obey \cite{Regge:1957td, Zerilli:1970wzz, Nichols:2012jn, Li:2023ulk}, 
\begin{align}
    \hat{\text{P}} h^{\rm (A)}_{\mu \nu} & = (-1)^{\ell+1} h^{\rm (A)}_{\mu \nu}, \\
    \hat{\text{P}} h^{\rm (P)}_{\mu \nu} & = (-1)^{\ell} h^{\rm (P)}_{\mu \nu}. 
\end{align}
Here $\hat{\text{P}}$ is the parity reversal operator, 
\begin{align}\label{eq:similarity}
    \hat{\text{P}} f(t, r, \chi, \phi) = f(t, r, -\chi, \pi + \phi), 
\end{align}
where $\chi = \cos \theta$. 

In the wave zone, the axial and polar perturbations map to the two GW waveform polarizations $h^+_{\mu\nu}$ and $h^\times_{\mu\nu}$,
\begin{align}
    \begin{pmatrix}h^{\rm (A)}\\h^{\rm (P)}\end{pmatrix} & = \begin{pmatrix}T_{\rm A, +} & T_{\rm A,\times} \\ T_{\rm P, +} & T_{\rm P,\times} \end{pmatrix}\begin{pmatrix}h^{+}\\h^{\times}\end{pmatrix},
\end{align}
where the $T_{\rm A/P, +/\times}$ are projection coefficients which depend on the orientation of the source.

For vacuum BHs in GR, the axial and the polar perturbations share the same QNM frequencies so for a particular set of QNM indices $(\ell,m,n)$\footnote{Only prograde QNMs are considered here, so we omit the final index $p\in\{+,-\}$.} all these metric terms oscillated with the same frequency, $\sim\exp(-i\omega_{lmn}t)$.
However, when isospectrality is broken, the axial and polar perturbations oscillate at different frequencies, $h^{\rm (A/P)}\sim e^{-i\omega_{lmn,\mathrm{A/P}}t}$, and the waveform contains a sum of both frequencies;
\begin{align}
    \begin{pmatrix}h^{+}\\h^{\times}\end{pmatrix} &= \begin{pmatrix}C_{+, \rm A} & C_{+, \rm P} \\ C_{\times, \rm A} & C_{\times, \rm P} \end{pmatrix}\begin{pmatrix}e^{-i\omega_{lmn,\mathrm{A}}t}\\e^{-i\omega_{lmn,\mathrm{P}}t}\end{pmatrix},
\end{align}
where the $C_{\rm +/\times, A/P}$ amplitude coefficients depend both on the amplitudes of the respective perturbations and on the orientation of the source.

We construct a phenomenological waveform to model this isospectrality breaking based on \texttt{pSEOBNR}. 
This is a multipolar, spin-precessing model for quasicircular inspirals, based on the effective-one-body formalism\footnote{Here, the $(\ell, |m|) =$ (2,2), (2,1), (3,3), (3,2), (4,4), (4,3), modes were included.}.
A key feature of this model is that its fundamental-mode ringdown frequency can be freely parameterized. 
The \texttt{pSEOBNR} model has been used extensively for BH spectroscopy and tests of GR \cite{LIGOScientific:2021sio, Pompili:2025cdc, LIGOScientific:2026wpt, LIGOScientific:2026oim}. 
We denote $h^{\rm (pSEOB)} (t; \bm{\Theta}, f_{l m 0}, \tau_{l m 0} )$ the time-domain \texttt{pSEOBNR} waveform model generated given the source parameters $\bm{\Theta}$ (such as initial masses and spins etc). 
With this notation, a GW strain that respects isospectrality can be written as \cite{Pompili:2025cdc}
\begin{align}
    h^{\rm (pSEOB)}(t) = \sum_{\ell m} h_{\ell m}^{\rm (pSEOB)} (t; \bm{\Theta}, f_{l m 0}, \tau_{l m 0}) {}_{-2}Y_{\ell m}(\varphi, \iota). 
\end{align}
Here $f_{l m 0}$ and $\tau_{l m 0}$ are the real frequency and damping time of the $l m 0$-th QNM. 
In GR, $f_{l m 0}$ and $\tau_{l m 0}$ are completely and uniquely determined by $\bm{\Theta}$. 

To model broken isospectrality, we use the following mode-by-mode modification of \texttt{pSEOBNR}, 
\begin{align}
    h_{\ell m}(t) = \left(1-\delta A_{\ell m}\right) & h_{\ell m} (t; f^{(1)}_{\ell m}, \tau^{(1)}_{\ell m}) \\
     + \delta A_{\ell m} &h_{\ell m}(t; f^{(2)}_{\ell m}, \tau^{(2)}_{\ell m}) . \nonumber
\end{align} 
This equation is an elaborated version of Eq.~\eqref{eq:waveform_model}. 
Here $h_{\ell m}(t; f^{(i)}_{\ell m}, \tau^{(i)}_{\ell m})$ denotes the usual \texttt{pSEOBNR} evaluated with the ringdown frequency and damping time modified from the GR values,
\begin{align}\label{eq:relative_deviation}
    f^{(1, 2)}_{l m 0} & = f^{\rm GR}_{l m 0} \left( 1 + \delta f^{(1, 2)}_{l m 0} \right), \\
    \tau^{(1, 2)}_{l m 0} & = \tau^{\rm GR}_{l m 0} \left( 1 + \delta \tau^{(1, 2)}_{l m 0} \right). \nonumber
\end{align}
We emphasize again that the labels (1) and (2) serve only to distinguish the two branches and should not be interpreted as axial and polar parity sectors, nor as the plus and cross GW polarizations. 
We define branch 1 to be amplitude dominant in order to remove the label-switching degeneracy. 
The model-agnostic observables are therefore the frequency and damping-time splittings between the two branches. 
Their identification with axial- and polar-led modes requires a theory-specific mapping.

For each pSEOBNR branch, we adopt the convention that $t=0$ corresponds to the peak of the coprecessing-frame $(2,2)$-mode amplitude, which coincides with the common merger–ringdown attachment time. 
Since the two branches share identical inspiral–merger dynamics and differ only in the frequency and damping time of the selected QNM, this prescription places them on a common time origin.
The pSEOBNR polarizations returned on their native finite time supports are resampled onto a common uniformly sampled cyclic time grid, with $t=0$ defined at the peak of the coprecessing-frame $(2,2)$-mode amplitude. 
The waveforms are zero-padded outside their native support before the two branches are linearly combined. 
Note that this resampling is solely a numerical grid conversion; no additional peak-time or phase alignment is performed.
Since only the $(2,2,0)$ mode is varied in this analysis, from now on we shall omit the mode indices. 

\section{Real mixing at leading order}

Our phenomenological waveform model, Eq.~\eqref{eq:waveform_model}, uses a real parameter $\delta A$ to describe the relative contribution of the two QNM branches. 
QNM excitation coefficients are generally complex, so this restriction does not follow from isospectrality alone.
We show below that, if the two branches have the same excitation phase in the GR limit, complex corrections to their relative excitation enter only beyond leading order.

Let
\begin{equation}
\zeta=\left(\frac{\ell}{M}\right)^k, 
\end{equation}
where $k\geq4$, and $\zeta$ is the dimensionless coupling parameter that controls the departure from GR.

We denote by $u_{\rm A}(t|\zeta)$ and $u_{\rm P}(t|\zeta)$ the time dependence of the axial and polar QNM sectors. 
We choose $t=0$ as a common reference time and fix their complex normalization by
\begin{equation}
u_{\rm A}(0|\zeta)=u_{\rm P}(0|\zeta)=1.
\label{eq:normalization}
\end{equation}
This fixes both the magnitude and phase of each function at $t=0$.
The excitation amplitudes and phases are captured by the complex coefficients $\mathcal{C}_{\rm A/P}$. 
Note that Eq.~\eqref{eq:normalization} is not a normalization with respect to a detector noise-weighted inner product.

For the perturbative theories considered here, the two normalized
functions reduce to the same function in the GR limit:
\begin{align}
u_{\rm A}(t|\zeta)
&=
u_{\rm GR}(t)
+\zeta u_{\rm A}^{(1)}(t)
+\mathcal{O}(\zeta^2),\\
u_{\rm P}(t|\zeta)
&=
u_{\rm GR}(t)
+\zeta u_{\rm P}^{(1)}(t)
+\mathcal{O}(\zeta^2).
\end{align}
Therefore,
\begin{equation}
u_{\rm P}(t|\zeta)-u_{\rm A}(t|\zeta)
=
\zeta
\left[
u_{\rm P}^{(1)}(t)-u_{\rm A}^{(1)}(t)
\right]
+\mathcal{O}(\zeta^2).
\label{eq:shape_difference}
\end{equation}

The ringdown waveform can be written as
\begin{equation}
h_{\rm RD}(t|\zeta)
=
\mathcal{C}_{\rm A}(\zeta)u_{\rm A}(t|\zeta)
+
\mathcal{C}_{\rm P}(\zeta)u_{\rm P}(t|\zeta),
\label{eq:branch_decomposition}
\end{equation}
where $\mathcal{C}_{\rm A/P} (\zeta)$ are complex in general. 
Defining
\begin{equation}
\mathcal{C} (\zeta) = \mathcal{C}_{\rm A} (\zeta) + \mathcal{C}_{\rm P} (\zeta),~ \text{and}~
\alpha (\zeta) = \frac{\mathcal{C}_{\rm P} (\zeta)}{\mathcal{C} (\zeta)}
\end{equation}
gives
\begin{equation}\label{eq:general_complex_mixing}
\begin{split}
h_{\rm RD}(t|\zeta) = \mathcal{C}(\zeta) \big[ & (1-\alpha(\zeta))u_{\rm A}(t|\zeta) \\
& +\alpha(\zeta) u_{\rm P}(t|\zeta) \big].
\end{split}
\end{equation}

When $\zeta=0$, $u_{\rm A}=u_{\rm P}=u_{\rm GR}$, and hence
\begin{equation}\label{eq:RD_GR}
h_{\rm RD}(t|0)= \mathcal{C}(0)u_{\rm GR}(t),
\end{equation}
independently of $\alpha(0)$. 
Thus, the GR waveform does not itself allow unique determination of $\mathcal{C}_{\rm P/A}$, and the total complex excitation ($\mathcal{C} (0)$) is observable.

We assume that the two branches have the same excitation phase at
$\zeta=0$. Their leading coefficients can then be written as
\begin{equation}\label{eq:real_leading_coefficients}
\begin{split}
\mathcal{C}_{\rm A}(0) & = (1-\delta A)\mathcal{C}(0), \\
\mathcal{C}_{\rm P}(0) & = \delta A\,\mathcal{C}(0),
\end{split}
\end{equation}
where $\delta A\in\mathbb{R}$. 
This is an assumption of our waveform model, rather than a consequence of isospectrality.

Allowing a general complex correction, we write
\begin{equation}
\alpha(\zeta)
=
\delta A
+
\zeta\alpha_1
+
\mathcal{O}(\zeta^2),
\label{eq:alpha_expansion}
\end{equation}
where $\alpha_1$ may be complex. Substituting the above expansions into
Eq.~\eqref{eq:general_complex_mixing} gives
\begin{equation}
\frac{h_{\rm RD}}{\mathcal{C}}
=
u_{\rm GR}
+
\zeta
\left[
(1-\delta A)u_{\rm A}^{(1)}
+
\delta A\,u_{\rm P}^{(1)}
\right]
+
\mathcal{O}(\zeta^2).
\label{eq:leading_real_mixing}
\end{equation}
The correction proportional to $\alpha_1$ starts at second order because
\begin{align}
&[\alpha(\zeta)-\delta A]
[u_{\rm P}(t|\zeta)-u_{\rm A}(t|\zeta)]
\nonumber\\
&\qquad
=
\zeta^2\alpha_1
\left[
u_{\rm P}^{(1)}(t)-u_{\rm A}^{(1)}(t)
\right]
+
\mathcal{O}(\zeta^3).
\end{align}
Thus, under the assumption in
Eq.~\eqref{eq:real_leading_coefficients}, complex corrections to the
relative excitation affect the waveform only beyond leading order.

A relative excitation phase that remains finite as $\zeta\rightarrow0$
would instead make $\alpha(0)$ complex and would affect the waveform at
$\mathcal{O}(\zeta)$. Such a phase is an additional leading-order
parameter that is not included in the present model. Under the
assumption made here, a real $\delta A$ is therefore sufficient at
leading order.

In our analysis, the two phenomenological branches are labelled $1$
and $2$, with branch $1$ defined to have the larger amplitude. This
gives $0\leq\delta A\leq1/2$. These labels are not fixed
identifications with the axial and polar sectors.

\section{Bayesian inference} \label{sec:Bayesian}

The phenomenological waveform model that we constructed enables us to compare via Bayesian inference the two hypotheses defined in the main text:
\begin{itemize}
    \item $H_{I}$: the null hypothesis, GR is correct and the signal respects isospectrality, modeled using $\underline{h}$ with the parameters $\bm{\Theta}$,
    \item $H_{\barredI}$: the alternative hypothesis, the signal breaks isospectrality, modeled using Eq.~\ref{eq:waveform_model} with the additional parameters $\delta A$, $\delta f^{(1,2)}$, and $\delta \tau^{(1,2)}$.
\end{itemize}

Testing these hypotheses amounts to computing their evidence, which involves the integration of the posterior probability distribution (henceforth, "posterior") of various parameters. 
The posterior of various parameters, given the data $d$, the hypothesis $H_{\barredI}$, and the background information $I$, can be expressed as follows according to the Bayes theorem, 
\begin{align}
    & p(\bm{\delta}, \delta A, \bm{\Theta}| d, H_{\barredI}, I) \\
    & = \frac{p(d |\bm{\delta}, \delta A, \bm{\Theta}, H_{\barredI}, I)p(\bm{\delta}, \delta A,  \bm{\Theta}| H_{\barredI}, I)}{p(d|H_{\barredI}, I)}, \nonumber
\end{align}
where $\bm{\delta} = \{\delta f^{(1)}, \delta f^{(2)}, \delta \tau^{(1)}, \delta \tau^{(2)}\}$, $p(\bm{\delta}, \delta A, \bm{\Theta}| d, H_{\barredI}, I)$, $p(d |\bm{\delta}, \delta A, \bm{\Theta}, H_{\barredI}, I)$, and $p(\bm{\delta}, \delta A, \bm{\Theta}| H_{\barredI}, I)$, are the posterior, likelihood, and prior probability distribution functions respectively. 
$p(d|H_{\barredI}, I)$ is the evidence of $H_{\barredI}$. 
Up to an additive constant, the likelihood is given by \cite{Thrane:2018qnx}
\begin{align}\label{eq:loglikelihood}
    \log p(d |\bm{\delta}, \delta A, \bm{\Theta}, H_{\barredI}, I) = - \frac{1}{2}\| d - h(\bm{\delta}, \delta A, \bm{\Theta})\|^2,  
\end{align}
where $\|a\|^2 = \sum_J \braket{a}_J$, and the bra-ket notation denotes the noise-weighted inner product, 
\begin{align}\label{eq:noise_weighted_inner_product}
    \langle a|b\rangle_J = 4 \text{Re} \int\text{d}f~ \frac{\tilde{a}(f) \tilde{b}^{\dagger}(f)}{S^{(J)}_n (f)}, 
\end{align}
$\tilde{a}(f)$ and $\tilde{b}(f)$ denote the Fourier transform of $a(t)$ and $b(t)$ respectively, the dagger sign $\dagger$ denotes complex conjugation, and $S^{(J)}_n (f)$ is the one-sided power spectral density of the detector $J$. 
The marginalized posterior is given by 
\begin{align}
    p\left(\bm{\delta}| d, H_{\barredI}, I\right) = \int d \delta A \int d \bm{\Theta} ~p\left(\bm{\delta}, \delta A, \bm{\Theta}| d, H_{\barredI}, I\right). 
\end{align}
\textit{If} the results of individual events $d = \{d_k|k = 1, 2, 3, \dots, N_{\rm event}\}$ are consistent with isospectrality, and a uniform prior is prescribed over $\bm{\delta}$, the marginalized posterior of $\bm{\delta}$ can be combined across the events, 
\begin{align}\label{eq:PDF_combination}
    p\left(\bm{\delta}| \{d \}, H_{\barredI}, I \right) \propto \prod_{k=1}^{N_{\rm event}} p\left(\bm{\delta}| d_k, H_{\barredI}, I \right). 
\end{align}

Table \ref{tab:priors} list the prior which we prescribe on various parameters for parameter estimation. 
When analyzing the actual GW signals, the upper and lower limits of the prior of chirp mass and luminosity distance are specified according to the results of parameter estimations of the signal using a waveform model consistent with GR (see Tab.~\ref{tab:priors}). 
Note that we place a prior which is linear in $\delta A$, because $\delta A$ is unlikely to occur unless the orientation of the binary is fine tuned. 
More often than not, $\delta A > 0 $ if the binary is viewed from a random orientation. 
Following \cite{Pompili:2025cdc}, we analyze both the mock and actual signals using the nested sampler \texttt{dynesty} \cite{Speagle:2019ivv} wrapped in the \texttt{bilby} package \cite{bilby_paper}, using the random-walk scheme (\texttt{acceptance-walk} stepping scheme). 
Following \cite{Pompili:2025cdc} , we sample the posterior distributions using 1000 live points (\texttt{nlive=1000}) and set the number of accepted MCMC-chains to be 60 (\texttt{naccept=60}). 

The Bayes factor can then be computed as 
\begin{align} \label{eq:B_factor}
    \mathcal{B}^{\barredI}_{I} = \frac{p(d|H_{\barredI}, I)}{p(d|H_{I}, I)},  
\end{align}
where the evidences $p(d|H, I)$, for $H = H_I $ or $H_{\barredI}$, is given by 
\begin{align}
    & p(d|H, I) \\
    & = \int d \bm{\delta} \int d \delta A \int d \bm{\Theta} ~p(d |\bm{\delta}, \delta A, \bm{\Theta}, H, I)p(\bm{\delta}, \delta A,  \bm{\Theta}| H, I). \nonumber
\end{align}
If $\mathcal{B}^{\barredI}_{I} <1$, $H_I$ is favored by the data, and $H_{\barredI}$ otherwise. 
The Bayes factor across $N_{\rm event}$ different events can be combined as a serial product, 
\begin{align}
    \mathcal{B}^{\barredI}_{I}  = \prod_{k=1}^{N_{\rm event}} \mathcal{B}^{\barredI, k}_{I} = \prod_{k=1}^{N_{\rm event}} \frac{p(d_k |H_{\barredI}, I)}{p(d_k|H_{I}, I)}. 
\end{align}

\begin{figure}
    \centering
    \includegraphics[width=\linewidth]{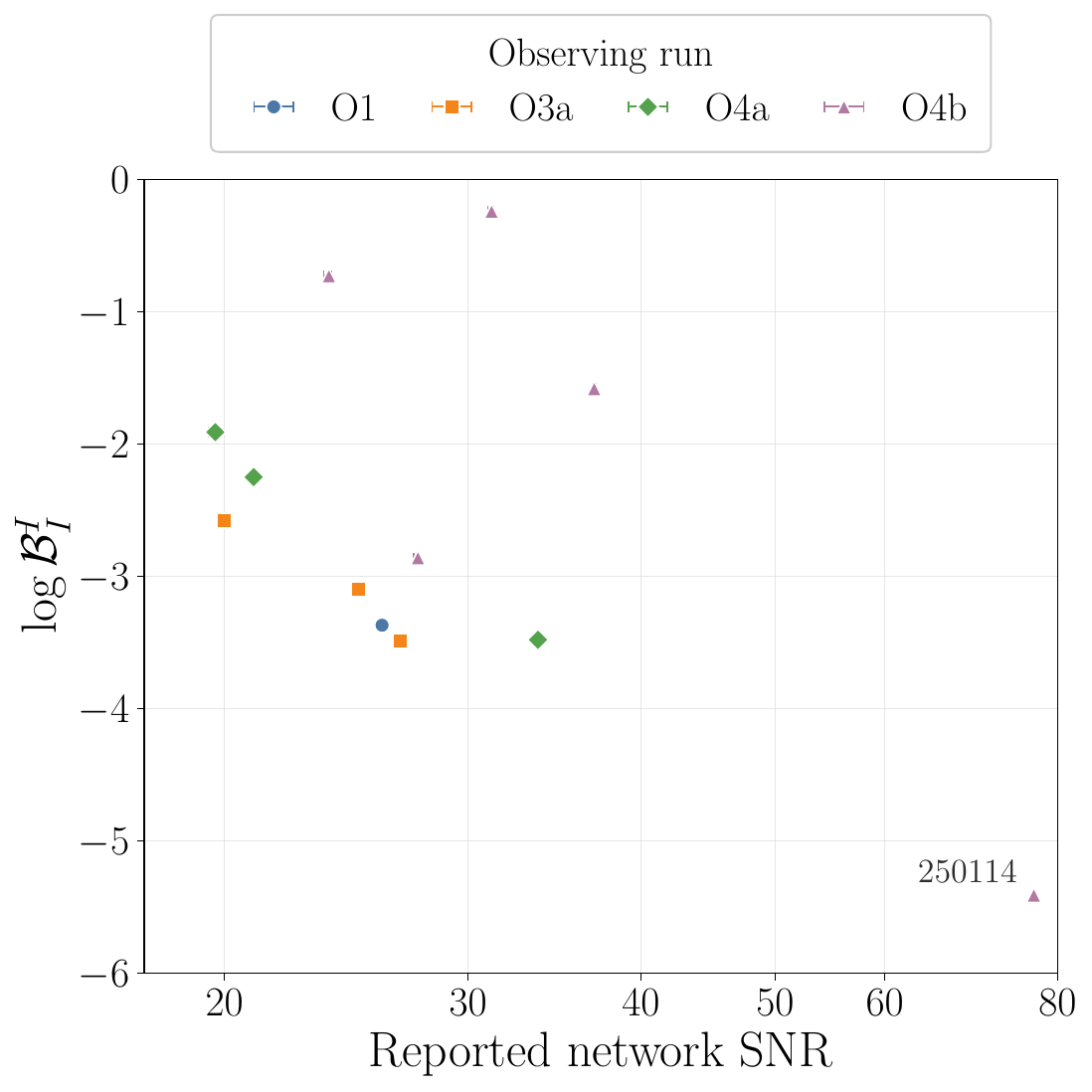}
    \caption{ \label{fig:Bayes_factor}
        Natural logarithm of the Bayes factor, $\log \mathcal{B}^{\barredI}_{I}$ (see Eq.~\eqref{eq:B_factor} in main text for definition), as a function of the reported network SNR for the 12 events analyzed in this work, with GW231028\_153006 excluded. 
        Positive values favor the non-isospectral hypothesis $H_{\barredI}$, whereas negative values favor the isospectral hypothesis $H_I$; the horizontal line at zero denotes equal evidence for the two hypotheses. 
        Colors and marker styles identify the observing runs, and the horizontal error bars indicate the reported 90\% credible intervals on the network SNR. 
        GW250114\_082203 is labeled as the event providing the strongest support for $H_I$.
    }
\end{figure}

\section{Model parameters and priors} \label{supp:priors}

The full list of model parameters and their priors is given in Table \ref{tab:priors}.
For the new, isospectrality-breaking frequency parameters uninformative uniform priors were used.
For the isospectrality-breaking amplitude parameter, $\delta A$, a linear prior was used. 
This was done to model the expected projection of the axial and polar perturbations on to the GW waveform.
The observed mix of axial and polar perturbations is controlled by various angles (including the viewing direction) that control how each perturbation projects on to the two GW polarisation states. 
Seeing only one type of perturbation (axial, say) requires a fine-tuning of these angles and is therefore unlikely \emph{a priori}. 
A roughly equal mix of the two perturbation types is more likely.

\begin{table*}[t]
    \begin{ruledtabular}
        \begin{tabular}{llr}
            Parameter type & Parameter & Prior\\
            \hline
            masses & chirp mass (detector frame) & linear, $\pi(\mathcal{M})\propto\mathcal{M}$ for $\mathcal{M}$ in event dependent range\\
            &mass ratio & $\pi(q)\propto (1+q)^{2/5}q^{-6/5}$ for $q\in[1/5, 1]$\\
            \hline
            spins & spin magnitudes &  uniform, $\chi_1,\chi_2\in [0, 0.99]$\\
            &spin tilts &  uniform in cosine, $\cos\theta_{1},\cos\theta_{2}\in [-1,1]$\\
            &spin separation angle &  periodic, uniform, $\phi_{12}\in [0, 2\pi]$\\
            &azimuthal precession angle & periodic, uniform, $\phi_{\rm JL}\in [0, 2\pi]$\\
            \hline
            extrinsic &right ascension &  periodic, uniform, $\alpha\in [0,2\pi)$\\
            &declination &  uniform in sine, $\sin\delta\in [-1,1]$\\
            & inclination &  uniform in cosine, $\cos\theta_{\rm JN}\in [-1,1]$\\
            &luminosity distance & uniform in co-moving volume,
            event-dependent bounds\\
            &coalescence time & uniform, for $t_{\rm c}$ in $t_c-t_{\rm ref}^{(i)}\in[-0.1,0.1]\,{\rm s}$\\
            &polarisation & periodic, uniform, $\psi\in 
            [0,\pi)$\\
            &phase & periodic, uniform, $\phi\in [0,2\pi)$\\
            \hline
            isospectrality breaking & amplitude &  linear, $\pi(\delta A)\propto\delta A$ for $\delta A\in [0,1/2]$ \\
            &frequency shifts  &  uniform, $\delta f^{(1)},\delta f^{(2)}\in [-4/5,1]$ \\
            &damping time shifts  &  uniform, $\delta \tau^{(1)},\delta \tau^{(2)}\in [-4/5,1]$\\
        \end{tabular}
    \end{ruledtabular}
    \caption{ \label{tab:priors}
        Model parameters and prior distributions.
        All the parameters listed here are independent; i.e.\ the priors are multiplied.
        The sampling was performed on the detector-frame chirp mass and the mass ratio using the priors induced from uniform distributions on the detector-frame component masses. 
        The luminosity-distance prior was uniform in comoving volume assuming the Planck15 flat-\(\Lambda\)CDM cosmology \cite{Planck:2015fie}.
        The parameters are broken down by type with the five new, beyond GR parameters shown at the bottom.
    }
\end{table*}

\section{Analysis of actual signals} \label{sec:actual_signals}

As mentioned in the main text, not all GW signals are equally informative for tests of isospectrality.
Signals with insufficient signal-to-noise ratio (SNR) generally provide weak constraints on the additional waveform degrees of freedom and may yield posteriors that are largely prior dominated.
We therefore restrict our analysis to a subset of the binary-BH mergers reported in GWTC-5.0, covering observing runs O1--O4b.
Specifically, we consider only events that satisfy the selection criteria adopted for the pSEOBNR analysis in Ref.~\cite{LIGOScientific:2026oim}.
These criteria require an SNR of $\geq 8$ in both the pre- and post-inspiral portions of the signal, which helps break the degeneracy between the fundamental ringdown frequency and the remnant mass.
In addition, we impose a threshold of $\mathrm{SNR}\geq18$ on the total signal to focus on the loudest and most informative events, which are expected to provide the strongest constraints on isospectrality breaking.

Among the events satisfying these criteria, we exclude GW231028\_153006 from our primary analysis.
This event is known to exhibit significant waveform-dependent systematic variations in its inferred source parameters, particularly the component masses \cite{LIGOScientific:2025slb}.
Furthermore, its inferred remnant spin lies outside the region over which QNM frequencies are currently available for several of the modified-gravity theories considered in our theory-specific interpretation.
Since these QNM calculations are required to map the model-independent constraints onto fundamental-physics parameters, we do not include GW231028\_153006 in the combined analysis.
The resulting sample consists of the 12 events listed in Tab.~\ref{tab:GW_events}.

The detector strain data used in our analysis are publicly available through GWOSC and were recorded by the Advanced LIGO~\cite{LIGOScientific:2014pky}, Advanced Virgo~\cite{VIRGO:2014yos}, and KAGRA~\cite{KAGRA:2020tym} detectors.
All events were analyzed using a sampling frequency of 2048 Hz, except GW230927\_153832, for which we used 4096 Hz. 
Owing to the comparatively low mass of this system, its higher-order $(4,4)$ merger–ringdown mode extends to higher frequencies and is not safely contained within the Nyquist frequency of 1024 Hz associated with a 2048 Hz sampling rate. 
The higher sampling frequency raises the Nyquist frequency to 2048 Hz, allowing the waveform to be represented without high-frequency truncation or aliasing. 
The likelihood was nevertheless evaluated over the same $20 - 1024 ~ \mathrm{Hz}$ frequency range used for all other events; thus, this exception changes only the waveform and strain sampling, rather than the frequency bandwidth entering the inference.

We sample the detector-frame chirp mass, $\mathcal{M}$, and mass ratio, $q=m_2/m_1\leq1$, using the corresponding \texttt{bilby} priors whose joint Jacobian yields a distribution uniform in the detector-frame component masses. 
To improve sampling efficiency while retaining the full likelihood support, the mass ranges are chosen separately for each event. 
Specifically, we determine the 0.1\% and 99.9\% quantiles of $\mathcal{M}$ from the publicly released posterior samples obtained with full inspiral-merger-ringdown GR models, enlarge this interval on each side by 25\% of its width, and round the resulting boundaries outwards. 
The prior on $q$, together with the constraints on the derived component masses $m_1$ and $m_2$, is chosen sufficiently broadly to contain the corresponding public posterior support. 
We verify using preliminary GR analyses that the recovered mass posteriors do not rail against any prior boundary and that all prior-supported configurations satisfy the waveform-generation validity requirements. 
For each event, identical astrophysical mass priors are employed in the GR and broken-isospectrality analyses. 
We additionally confirm on representative events that modest broadening of the mass ranges does not materially affect the inferred isospectrality-breaking parameters or the corresponding Bayes factors.

The luminosity-distance prior was taken to be uniform in comoving volume and source-frame time, with event-dependent bounds chosen from the publicly released LVK posterior samples
\footnote{Specifically, we used the cosmologically reweighted \texttt{C01:Mixed} samples from the GWTC-2.1 and GWTC-3
parameter-estimation data releases~\cite{LVK_GWTC21_PE,LVK_GWTC3_PE}, the \texttt{C00:Mixed} samples from the GWTC-4.1
release~\cite{LVK_GWTC41_PE}, and, for the GWTC-5.0/O4b events, the \texttt{C00:NRSur7dq4} samples~\cite{LVK_GWTC50_PE_I,LVK_GWTC50_PE_II}, since mixed-waveform posterior samples were not publicly available for those events.}. 
For each event, we first determined the (0.1\%) and (99.9\%) posterior quantiles, denoted by $L_{d_L}$ and $U_{d_L}$, respectively, and extended the interval on each side by (25\%) of its width:
\begin{align}
    d_{L,\min} & =L_{d_L}-0.25\left(U_{d_L}-L_{d_L}\right),\\
    d_{L,\max} & =U_{d_L}+0.25\left(U_{d_L}-L_{d_L}\right).
\end{align}
The resulting bounds were rounded outwards to the nearest 100 Mpc, with a minimum lower bound of 100 Mpc. 
This procedure provides sufficiently broad event-specific sampling ranges without adopting the LVK posterior distribution itself as the prior; the functional form of the luminosity-distance prior remains identical for all events.

For each of these events, we estimated the power spectral density [$S_n(f)$ in Eq.~\eqref{eq:noise_weighted_inner_product}] as follows. 
We first select 128 seconds of data before and after the 4 seconds of data which encloses the signals. 
The average power spectral density of these two segments of 128-second data is used for the evaluation of likelihood [c.f. Eq~(\ref{eq:loglikelihood})]. 

Since the marginalized posterior of $\delta A$ skews toward 0.5, the upper limit of its range covered by its prior, we quote the 90-\% upper quantile of the marginalized posterior of $\delta A$, $\delta A^{(90)}$, defined as 
\begin{align}
    \int_{0}^{\delta A^{(90)}} d \delta A~p(\delta A|d, H_{\barredI}, I) = 0.9
\end{align}

Finally, to compute the Bayes factor [c.f. Eq.~\eqref{eq:B_factor}], for each of these events, we also perform a reference inference in which $\delta A, \delta f^{(1,2)}$ and $\delta \tau^{(1,2)}$ are pinned at zero, so that our waveform model effectively reduces to the original \texttt{pSEOBNRv5PHM} waveform model.

\section{Deriving constraints on gravity theories} \label{sec:physics_implications}

Although the parameterization of our phenomenological waveform model in terms of $\bm{\delta}$ is convenient for parameter estimation, it is not the most optimal for interpreting the fundamental-physics implications of our analysis, such as constraining the coupling length scales of specific beyond-GR theories.
For this purpose, we introduce the alternative parameterization
\begin{align}
    \mu_{f} & = \frac{1}{2} \left( \delta f^{(1)}_{l m 0} + \delta f^{(2)}_{l m 0} \right), \label{eq:mu_f} \\
    \mu_{\tau} & = \frac{1}{2} \left( \delta \tau^{(1)}_{l m 0} + \delta \tau^{(2)}_{l m 0} \right), \label{eq:mu_tau} \\
    \Delta_{f} & = \left| \delta f^{(1)}_{l m 0} - \delta f^{(2)}_{l m 0} \right|, \label{eq:Delta_f} \\
    \Delta_{\tau} & = \left|  \delta \tau^{(1)}_{l m 0} - \delta \tau^{(2)}_{l m 0} \right| \label{eq:Delta_tau}.
\end{align}
The parameters $\mu_{f}$ and $\mu_{\tau}$ describe shifts common to the two QNM branches and can therefore be partially degenerate with changes in the inferred remnant mass and spin.
By contrast, $\Delta_{f}$ and $\Delta_{\tau}$ are label-invariant measures of the frequency and damping-time splittings between the two branches.
They therefore provide the natural model-independent observables for testing isospectrality and, through theory-specific mappings of the QNM spectrum, for constraining the coupling scales of beyond-GR theories that predict isospectrality breaking.

If the events are described by a common frequency and damping-time splitting, the posterior distributions of $\mu_f, \mu_{\tau}, \Delta_{f}$ and $\Delta_{\tau}$ can be combined across the signals in a manner analogous to
Eq.~\eqref{eq:PDF_combination}.
Special care is required, however, because the uniform priors on $\delta f^{(1,2)}$ and $\delta \tau^{(1,2)}$ do not induce uniform priors on the corresponding shift and splitting parameters. 
For two independent variables uniformly distributed over an interval of width $W$, the prior on their absolute difference is triangular.
Consequently,
\begin{equation}
\begin{split}
    p\left(\Delta_{f}, \Delta_{\tau}|H_{\barredI}, I \right) & = \frac{4}{W^4} \left(W-\Delta_{f} \right) \left(W-\Delta_{\tau} \right), \\
    p\left(\mu_f, \mu_{\tau}|\Delta_{f}, \Delta_{\tau}, H_{\barredI}, I \right) & = \frac{1}{\left(W-\Delta_{f} \right) \left(W-\Delta_{\tau} \right)}.
\end{split}
\end{equation}
Thus, the joint prior on $\mu_f, \mu_{\tau}, \Delta_f$ and $\Delta_{\tau}$ is 
\begin{equation}
p\left(\mu_f, \mu_{\tau}, \Delta_{f}, \Delta_{\tau}| H_{\barredI}, I \right) = \frac{4}{W^4}. 
\end{equation}
Here $W=1-(-0.8)=1.8$ is the width of the priors on $\delta f^{(1,2)}$ and $\delta \tau^{(1,2)}$. 
By Bayes' theorem, multiplying the event-level posteriors directly would count this prior once for each event. 
Assuming that $\mu_f, \mu_{\tau}, \Delta_f$ and $\Delta_{\tau}$ are the same across all the events, the combined posterior is
\begin{widetext}
\begin{align}
    p\left(\mu_f, \mu_{\tau}, \Delta_{f}, \Delta_{\tau}|\{d\}, H_{\barredI}, I \right) \propto \left[ p\left(\mu_f, \mu_{\tau}, \Delta_{f}, \Delta_{\tau}|H_{\barredI}, I \right) \right]^{1-N_{\rm event}} \prod_{k=1}^{N_{\rm event}} p\left(\mu_f, \mu_{\tau}, \Delta_{f}, \Delta_{\tau}|d_k, H_{\barredI}, I \right). 
\end{align}
\end{widetext}
Marginalizing the above posterior distribution over $\mu_f $ and $\mu_{\tau}$ yields the two-dimensional posterior of $(\Delta_f, \Delta_{\tau})$ shown in Fig.~\ref{fig:catalog_PDF}, 
\begin{equation}
\begin{split}
& p\left(\Delta_{f}, \Delta_{\tau}|\{d\}, H_{\barredI}, I \right) \\
& = \int d \mu_f \int d \mu_{\rm \tau}~p\left(\mu_f, \mu_{\tau}, \Delta_{f}, \Delta_{\tau}|\{d\}, H_{\barredI}, I \right). 
\end{split}
\end{equation}

In light of the non detection of isospectrality breaking, one can map the combined posterior of $\Delta_{f}$ and $\Delta_{\tau}$ into the coupling length scales of specific beyond-GR theories, using the QNM spectra therein. 
For a given theory and parity $X\in\{\mathrm{A},\mathrm{P}\}$, the leading-order complex QNM frequency can be written as Eq.~\ref{eq:qg_qnm_expansion} in the main text. 

Up to the leading order, the fractional frequency and damping-time shifts defined by Eq.~\eqref{eq:relative_deviation} are related to the QNM spectra in a specific gravity theory as 
\begin{align}
    \delta f_{\rm (X)}(\ell,M,a)
    &= \left( \frac{\ell}{M} \right)^k \frac{\operatorname{Re} \Omega_{\rm (X)} (a)} {\operatorname{Re} M \omega^{\rm GR}(a)},
    \label{eq:qg_delta_f}\\
    \delta\tau_{\rm (X)}(\ell,M,a)
    &= - \left( \frac{\ell}{M} \right)^k  \frac{\operatorname{Im} \Omega_{\rm (X)} (a)} {\operatorname{Im} M \omega^{\rm GR}(a)}.
    \label{eq:qg_delta_tau}
\end{align}
The Kerr frequency $\omega^{\rm GR}(a)$ is evaluated numerically.
For remnant spins above the calibrated range of the fitted QNM corrections, we extrapolate the corresponding spin polynomials up to $a_f=0.99$. 
Our results are insensitive to this extrapolation because the bulk of the posterior probability for the remnant spin magnitude lies around lower values of $\chi_f\sim 0.7$.
One of the present authors has previously investigated the impact of various extrapolations on the posterior on $\ell$, and found the effects negligible \cite{Chung:2025ucb}.
We emphasize that the resulting coupling-length distributions should therefore be regarded as effective leading-order mappings of the phenomenological posteriors, rather than direct theory-specific Bayesian posteriors.

For each event $k$, we construct a six-dimensional kernel-density estimate
(KDE) of
\begin{align}
    \mathbf{y}_k =\left(\delta f_k^{(1)},\delta\tau_k^{(1)}, \delta f_k^{(2)},\delta\tau_k^{(2)},a_{k},M_{k}\right)
\end{align}
from the equal-weight posterior samples, where we have omitted the mode label. 
When $M_k$ and $a_k$ are not stored explicitly, they are obtained sample by sample using the fit in the \texttt{NRSur7dq4Remnant} waveform model. 
All parameters not included in $\mathbf{y}_i$,
including the relative-amplitude parameter, are thereby marginalized.  
We average the two possible assignments of the phenomenological sectors to the axial and polar modes with equal prior weight.
Defining $\mathbf{d}_{(X)}=(\delta f_{(X)},\delta\tau_{(X)})$, the event-level likelihood-like curve is
\begin{align} \label{eq:event_theory_likelihood}
    & p(\ell|d_k, H_{\barredI}) \\
    & \propto \int  d M_k \int d a_k \left[ \widehat p_k \left(\mathbf{d}_{\rm A}^T,\mathbf{d}_{\rm P}^T,a_k,M_k | d_k , H_{\barredI} \right) + (\textbf{A} \leftrightarrow \textbf{P})\right], \nonumber
\end{align}
where, explicitly, 
\begin{align}
    & \widehat p_k \left(\mathbf{d}_{\rm A}^T, \mathbf{d}_{\rm P}^T, a_k, M_k | d_k , H_{\barredI} \right) \\
    & = \widehat p_k \big(\delta f_k^{(1)} = \delta f_k^{(A)}, \delta\tau_k^{(1)} = \delta\tau_k^{(A)}, \nonumber\\
    & \quad \quad \quad \delta f_k^{(2)} = \delta f_k^{(P)}, \delta\tau_k^{(2)} = \delta\tau_k^{(P)}, \nonumber\\
    & \quad \quad \quad a_{k},M_{k} | d_k , H_{\barredI} \big), \nonumber
\end{align}
and $(\textbf{A} \leftrightarrow \textbf{P})$ implies the interchange of parities $(A)$ and $(P)$.  
Note that Eq.~\eqref{eq:event_theory_likelihood} has already encoded the original priors on various parameters. 

As $\ell $ is supposed to be a constant, its posterior inferred from individual events can be combined to build a catalog posterior one, 
\begin{align}\label{eq:catalog_ell_posterior}
    p(\ell|\{d\},H_{\barredI}) \propto p(\ell|H_{\barredI})^{1-N_{\rm event}}\prod_{k=1}^{N_{\rm event}} p(\ell|d_k, H_{\barredI}),
\end{align}
where $p(\ell|H_{\barredI})$ is the prior on $\ell$ given $H_{\barredI}$, which we adopt to be a uniform prior on $\ell$ over the analysis range.

We apply this mapping to construct effective posteriors for the coupling length scale $\ell$ in several representative beyond-GR theories.
Specifically, we consider scalar Gauss-Bonnet, dynamical Chern-Simons, and axi-dilaton gravity, with the corresponding definitions of $\ell$ given in Ref.~\cite{Chung:2025ucb}, as well as cubic and quartic gravity theories, for which $\ell$ is defined through the Lagrangian density of Ref.~\cite{Maenaut:2024oci}, with $\lambda_2=0$.
For each theory, we use the fitting polynomials $\Omega_{\rm (X)}(a)$ obtained in Refs.~\cite{Chung:2024ira,Chung:2024vaf,Chung:2025gyg,Chung:2025ucb,Cano:2023jbk} to map the model-independent constraints on the QNM splittings onto $\ell$.

Figure \ref{fig:PDF_ell} shows the resulting effective posteriors for $\ell$.
For all theories considered, the posteriors are consistent with the GR limit, $\ell=0$, and exhibit a broadly similar shape: they remain approximately flat or slowly varying at small $\ell$ before decreasing rapidly beyond a characteristic scale set by the typical masses of the BHs.
This behavior is consistent with the absence of evidence for isospectrality breaking and therefore allows us to place upper bounds on the corresponding coupling scales.
We quote one-sided 90\% credible upper limits on $\ell$, which are summarized in Tab.~\ref{tab:theory_constraints}.
The resulting bounds are quantitatively consistent with those obtained from the theory-specific analyses of Refs.~\cite{Chung:2025ucb,Maenaut:2024oci}.
This agreement provides an independent consistency check of our mapping procedure and demonstrates that the model-independent isospectrality constraints derived here can be translated into meaningful bounds on the fundamental coupling scales of a broad class of modified-gravity theories.

\section{Analysis of mock signals} \label{sec:inj_study}

To better understand the results of our search in the absence of genuine isospectrality breaking, we analyze five mock signals that respect isospectrality.  
We generate a \texttt{pSEOBNR} waveform using the maximum-posterior source parameters inferred for GW250114~\cite{LIGOScientific:2025wao}, with $\delta f^{(1,2)}=\delta\tau^{(1,2)}=0$, and inject the same signal into several realizations of real detector noise.

For each of these realizations, the noise is constructed from 4-s off-source strain segments surrounding GW250114. 
For a Hanford segment beginning at GPS time $t_{\rm H}$, we use a Livingston segment beginning at $t_{\rm L}=t_{\rm H}+4\,{\rm s}$. 
Sliding the Livingston segments by 4 seconds suppresses any pre-existing inter-detector coherence in the off-source data while retaining the single-detector non-Gaussian character of real detector noise.
The resulting injections should therefore be interpreted as robustness tests against decorrelated realizations of real detector noise, rather than as reconstructions of a particular simultaneous network-noise epoch. 
The two noise segments are then retimed to a common analysis epoch before the same coherent mock signal is injected.  
Their network SNRs are comparable to that of GW250114. 
Each realization is analyzed under both $H_I$ and $H_{\bar I}$ to compute the corresponding Bayes factor.

Table~\ref{tab:mock_signal_log_B} summarizes the noise segments, GPS start time of the Hanford segments,  network SNRs, and corresponding Bayes factors.  
Although all five injected signals respect isospectrality, the recovered $\log\mathcal{B}^{\barredI}_{I}$ varies appreciably with the noise realization.  
In particular, Realization~4 gives $\log\mathcal{B}^{\barredI}_{I}=0.0463$, corresponding to essentially equal support for the two hypotheses, while Realization~5 gives only mild support for $H_I$, $\log\mathcal{B}^{\barredI}_{I}=-0.546$. 
The other realizations yield more negative Bayes factors. 
Thus, even for an exactly isospectral signal, real-noise fluctuations can move the model preference close to
either side of $\log\mathcal{B}^{\barredI}_{I}=0$. 
A Bayes factor slightly favoring the non-isospectral model should therefore not by itself be interpreted as evidence for black-hole fine structure.
This demonstrates that a slightly positive Bayes factor can arise from detector noise alone. 
It does not by itself account for the larger Bayes factor obtained for GW231028\_153006; rather, together with the known waveform-dependent systematics of that event, it motivates caution in interpreting such preferences.
A larger background or injection campaign would be required to calibrate the distribution of the Bayes factor under the isospectral hypothesis and to quantify how unusual a candidate value is.

\begin{table}[t]
    \begin{ruledtabular}
        \begin{tabular}{cccc}
            Realization
            & H1 GPS start time
            & SNR
            & $\log \mathcal{B}^{\barredI}_{I}$ \\
            \hline
            1
            & 1420878279
            & 78.1
            & -4.95 \\
            
            2
            & 1420878287
            & 75.4
            & $-5.73$\\
    
            3
            & 1420878295
            & 77.8
            & $-3.74$ \\
    
            4
            & 1420878303
            & 77.8
            & 0.0463\\
    
            5
            & 1420878311 
            & 77.9
            & $-0.546$\\
        \end{tabular}
    \end{ruledtabular}
    \caption{ \label{tab:mock_signal_log_B}
        Results of the isospectral injection study in different realizations of real detector noise. 
        The GPS time denotes the start time of the Hanford noise segment used for each realization, and the SNR is the network signal-to-noise ratio of the injected signal. 
        The final column gives the natural logarithm of the Bayes factor $\mathcal{B}^{\barredI}_{I}$, with positive (negative) values favoring the non-isospectral (isospectral) hypothesis.
    }
\end{table}

\begin{figure}
    \centering
    \includegraphics[width=\linewidth]{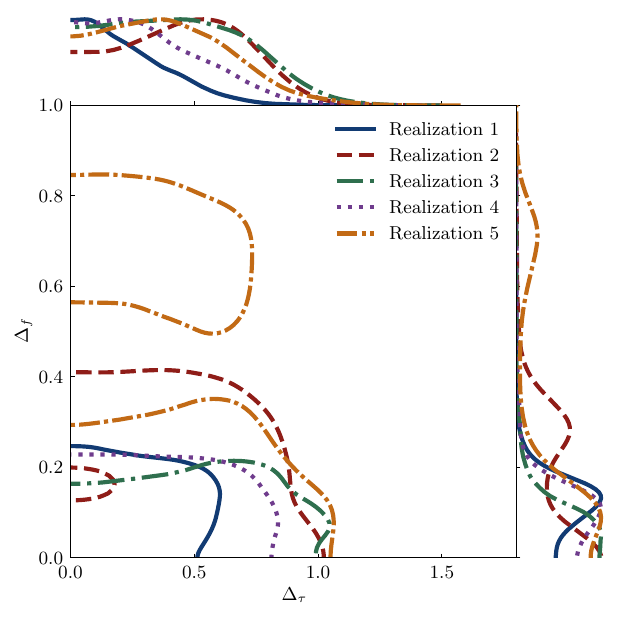}
    \caption{\label{fig:mock_injection_2D}
    Posterior distribution on the dimensionless frequency and damping-time splittings, $\Delta_f$ and $\Delta_\tau$, of the fundamental (2,2,0) QNM, obtained from five realizations of a GR mock injection in different segments of real detector noise. 
    The central panel shows the corresponding 90\% credible contours, while the upper and right panels show the normalized one-dimensional marginalized posterior distributions of $\Delta_\tau$ and $\Delta_f$, respectively. Each color and line style denotes an individual noise realization; the posteriors are not combined. 
    The GR isospectral limit corresponds to $(\Delta_\tau,\Delta_f)=(0,0)$. 
    Notably, Realization 5 exhibits an additional disconnected branch at large $\Delta_f$, illustrating that apparently anomalous event-level splitting posteriors can arise from particular noise realizations even when the injected signal satisfies isospectrality.
    }
\end{figure}

\begin{figure*}[htp!]
    \centering
    \includegraphics[width=\linewidth]{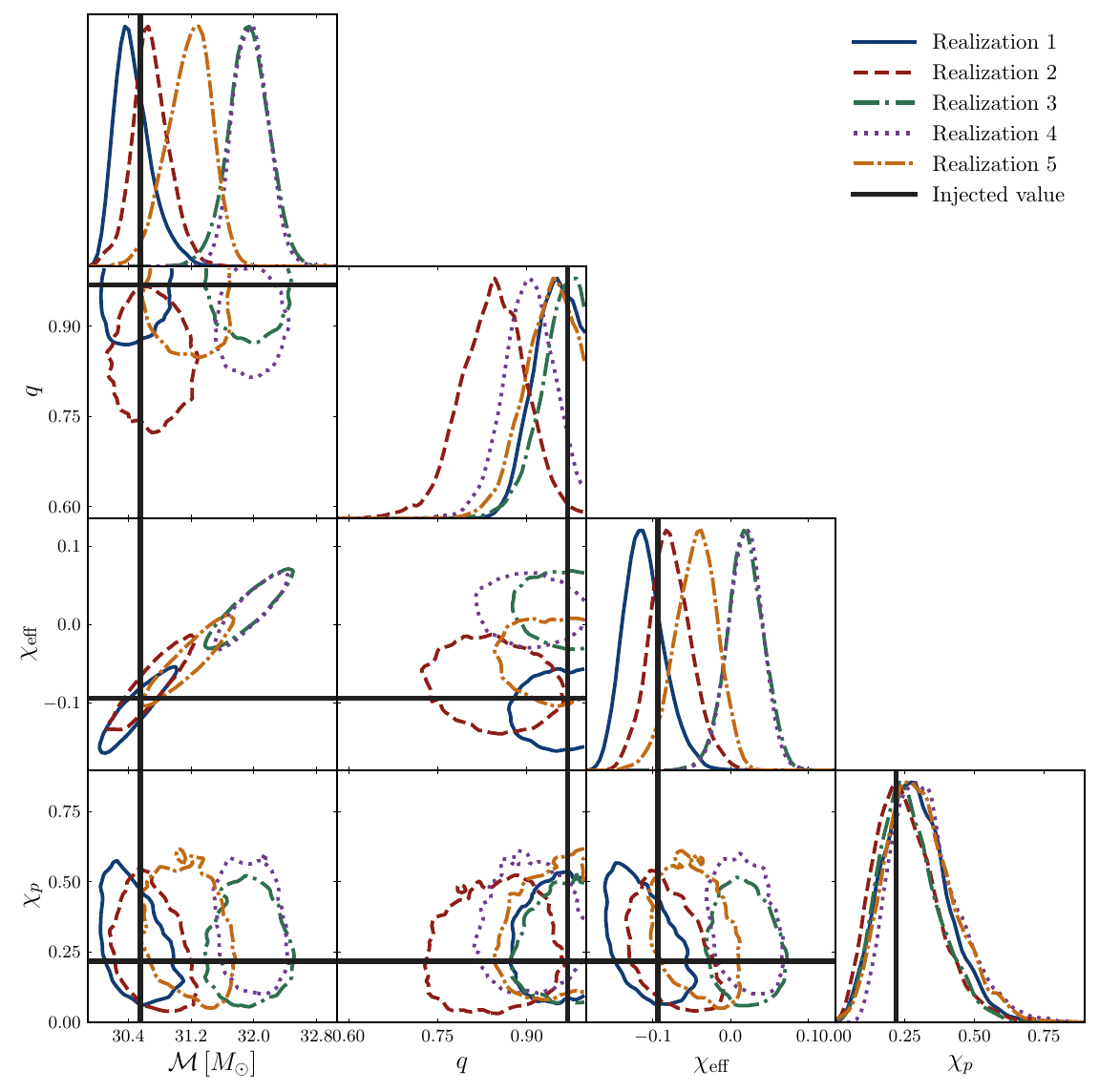}
    \caption{\label{fig:mock_corner_plot}
    Marginalized posterior distributions of the detector-frame chirp mass $\mathcal{M}$, mass ratio $q$, effective aligned-spin parameter $\chi_{\rm eff}$, and effective precession-spin parameter $\chi_p$ obtained from five realizations of the same GR mock signal injected into different 4-s segments of real detector noise. 
    Each mock signal is analyzed with the phenomenological waveform model defined in Eq.~\eqref{eq:waveform_model}, using the same setup as for the observed events.
    The diagonal panels show the one-dimensional marginalized posterior densities, normalized to unit peak, while the lower-triangle panels show the corresponding joint 90\% credible contours. 
    Colors and line styles distinguish the five noise realizations, and the thick black lines mark the common injected values. The posteriors from the individual realizations are shown separately and are not combined.   
    }
\end{figure*}

Figure~\ref{fig:mock_injection_2D} shows the corresponding posteriors on the splitting parameters $\Delta_f$ and $\Delta_\tau$.  
All five posteriors contain support near the isospectral limit ($\Delta_f = \Delta_{\tau} = 0$).
The first four are dominated by relatively small splittings, whereas Realization~5 develops an additional disconnected branch at large $\Delta_f$.  
Inspection of the signed-shift posterior shows that this feature originates from a secondary mode in $\delta f^{(2)}$, which is mapped to a large value of $\Delta_f$ by Eq.~\eqref{eq:Delta_f}. 
The appearance of this secondary branch in an exactly isospectral injection demonstrates that such posterior structure can be generated by the particular noise realization rather than by genuine isospectrality breaking.

Real detector noise also affects the estimation of source parameters. 
Fig.~\ref{fig:mock_corner_plot} shows the detector-frame chirp mass $\mathcal{M}$, mass ratio $q$, effective aligned and precession spins, ($\chi_{\rm eff}$ and $\chi_p$).
While Realizations 1, 2 and 5 result in posteriors that are largely consistent with the injected values, the posteriors of Realizations 3 and 4 show significant displacements in $\mathcal{M}$ and $\chi_{\rm eff}$. 
In both cases the posterior moves along a strong positive $\mathcal{M}$--$\chi_{\rm eff}$ correlation, rather than showing independent inconsistencies in the two parameters.
Such a mass--spin degeneracy is known to persist for high-mass binary-BH signals, for which the merger and ringdown contribute substantially to the measured waveform and the degeneracy is no longer well characterized by a constant chirp mass~\cite{Baird:2012cu, Roulet:2018jbe}.

\begin{figure}[htp!]
    \centering
    \includegraphics[width=\linewidth]{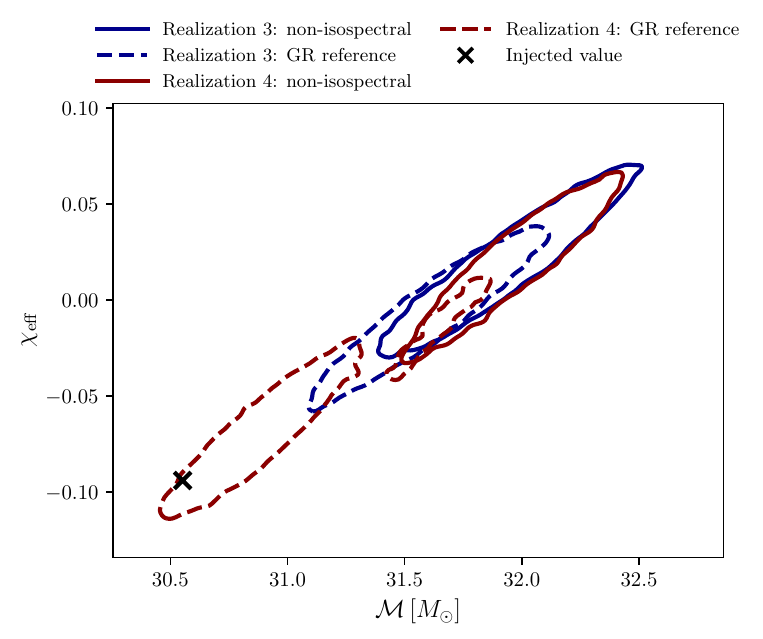}
    \caption{\label{fig:mock_mass_chieff_gr_compare}
    Comparison of the joint posterior distributions of the detector-frame chirp mass $\mathcal{M}$ and effective aligned-spin parameter $\chi_{\rm eff}$ for Realizations~3 and~4.
    Solid contours show the 90\% credible regions obtained with the non-isospectral waveform model, while dashed contours show the corresponding GR-reference recoveries with all QNM-deviation parameters fixed to zero.
    The cross marks the common injected values.
    For Realization~3, both analyses are displaced along approximately the  same $\mathcal{M}$--$\chi_{\rm eff}$ degeneracy. 
    For Realization~4, the GR-reference recovery remains considerably closer to the injected values, whereas the non-isospectral posterior extends much farther along the same intrinsic mass--spin degeneracy.
    }
\end{figure}

To investigate the origin of the comparatively large shifts in $\mathcal{M}$ and $\chi_{\rm eff}$, we compare the non-isospectral recoveries of Realizations~3 and~4 with corresponding GR-reference analyses in which all QNM-deviation parameters are fixed to zero.
The comparison is shown in Fig.~\ref{fig:mock_mass_chieff_gr_compare}.
In both realizations, the recovered parameters follow a strong positive $\mathcal{M}$--$\chi_{\rm eff}$ correlation, with the injected values lying approximately along the same degeneracy direction.
The two realizations nevertheless behave differently.
For Realization~3, the GR-reference posterior is already substantially displaced from the injected parameters along essentially the same $\mathcal{M}$--$\chi_{\rm eff}$ degeneracy as the non-isospectral recovery. 
A detector-by-detector likelihood decomposition shows that this displacement is driven predominantly by the time-slid Livingston noise realization. 
The shift is therefore already present in the standard GR parameter recovery and is not generated by the additional
QNM-deviation parameters.

For Realization~4, on the other hand, the GR-reference posterior remains substantially closer to the injected parameters, whereas the non-isospectral posterior moves much farther along the same $\mathcal{M}$--$\chi_{\rm eff}$ degeneracy.
In this realization, the additional QNM degrees of freedom provide greater flexibility to fit the particular detector-noise realization and allow the recovered source parameters to move farther along an existing intrinsic mass--spin degeneracy.
However, this behavior does not imply an equally strong degeneracy between the isospectrality-breaking observables $\Delta_f$ and $\Delta_\tau$ and the GR source parameters.
The individual QNM shifts also contain common-mode components, which can partially mimic changes in the remnant mass and spin, whereas the branch-to-branch splittings provide the direct observables of isospectrality breaking.
Furthermore, the corresponding improvement in maximum likelihood does not translate into significant Bayesian support for the non-isospectral hypothesis:
$\log\mathcal{B}^{\barredI}_{I}=0.0463$, which is consistent with essentially equal evidence for the two hypotheses.

These examples show that apparently large offsets in recovered source parameters need not represent independent inconsistencies in $\mathcal{M}$ and $\chi_{\rm eff}$. Real detector noise can instead displace the posterior coherently along a strong intrinsic waveform degeneracy, while additional testing degrees of freedom can, for some noise realizations, amplify the extent of this displacement without producing meaningful evidence for isospectrality breaking.

\end{document}